\documentclass[10pt,aps,twocolumn,showpacs,prl]{revtex4}
\usepackage{amssymb,amsfonts,amsmath,wrapfig,mathrsfs,mathbbol}
\usepackage{graphicx}
\begin{document}

\title{Measuring the Complete Transverse Spatial Mode Spectrum of a Wave Field}

\author{Gabriel F. Calvo,$^{1,2*}$ Antonio Pic\'{o}n,$^{1}$ and Roberta Zambrini$\,^{3}$}

\affiliation{$^{1}$Grup de F\'{\i}sica Te\`{o}rica, Universitat Aut\`{o}noma de Barcelona, 08193 Bellaterra (Barcelona), Spain}
\affiliation{$^{2}$Institute of Applied Mathematics in Science and Engineering, IMACI (UCLM), Ciudad Real, 13071, Spain}
\affiliation{$^{3}$Institute for Cross-Disciplinary Physics and Complex Systems, IFISC (CSIC-UIB), Palma de Mallorca, 07122, Spain}
\date{\today}

\begin{abstract}
We put forward a method that allows the experimental determination of  the entire spatial mode spectrum of any arbitrary monochromatic wave field in a plane normal to its propagation direction. For coherent optical fields, our spatial spectrum analyzer can be implemented with a small number of benchmark refractive elements embedded in a single Mach-Zehnder interferometer. We detail an efficient setup for measuring in the Hermite-Gaussian mode basis. Our scheme should also be feasible in the context of atom optics for analyzing the spatial profiles of macroscopic matter waves.  
\end{abstract}

\maketitle
The temporal~\cite{Ye05},  vectorial~\cite{Damask04} and spatial~\cite{Harrison04} degrees of freedom of electromagnetic fields with increasing complexity  are exploited in applications ranging from communications to medicine. The detailed knowledge of the {\em spatial mode spectrum} of these
beams, generally emitted by laser devices, is fundamental in optimizing applications as well as in research, both in classical and quantum optics. Many imaging and tailoring techniques, for instance, are based on the possibility to devise transformations induced by optical elements by determining the effects on the spatial components of generic signals~\cite{Goodman}. Recently, an intense research activity on multimode light has burgeoned in quantum information, communication and imaging~\cite{QuantImage}, with successful demonstrations spanning from nanodisplacement measurements~\cite{Treps03}, to parallel information~\cite{Lassen07}, high-dimensional entanglement~\cite{OAM}, coherent transfer of suitable prepared superpositions of vortices onto Bose-Einstein condensates~\cite{Andersen}, and hot Rubidium vapors~\cite{Pugatch}. A key ingredient in all these scenarios is {\em spatial encoding}, which conveys several independent channels of information by exploiting the access to a large number of optical transverse modes. The use of multimode beams thus demands for the development of techniques allowing to characterize their spatial spectrum and to retrieve the information encoded in different components. Fourier modes are of immediate access through a basic lens setup~\cite{Goodman}, while rotating elements lead to 'helical' spectra decomposition~\cite{Zambrini}. To the best of our knowledge, however, no {\em direct method} is known to measure the Hermite-Gaussian (HG) modes spectrum. Building on the symplectic formalism to describe first-order optical transformations, we present in this Letter a general strategy to find an arrangement of refractive elements that enables the quantitative measurement of the transverse spatial spectrum of light beams by using a single Mach-Zehnder interferometer. Due to the generality of our approach, different experimental setups can be implemented to extract the full spatial  spectrum of multimode transverse beams in many bases, including Laguerre-Gaussian (LG) modes. Here we focus on the spectrum analyzer for HG modes, as these are the most common in laser physics and appear naturally in devices where astigmatism, strain or slight misalignment drive the system toward rectangular symmetry~\cite{Siegman}. Furthermore, the presented framework is rather suggestive of analog measurements for matter waves spatial spectra.
\par
Spatial modes are ubiquitous. Sets of modes $u$ satisfying the wave equation $(i\partial_{\eta}+\partial^{2}_{x}+\partial^{2}_{y})u=0$ describe broad classes of physical systems. When the evolution variable is $\eta=2kz$, the sets comprise the optical paraxial modes~\cite{Siegman} with wave number $k$ propagating along the $z$ direction. For $\eta=2mt/\hbar$, the sets model instead the quantum dynamics of free particles of mass $m$ in the $xy$ plane. Since there are two transverse spatial variables, each entire set of mode solutions $u_{m,n}$ is labeled by two integers $m,n$. Relevant examples are the HG and LG mode bases. Any scalar field $\psi$ obeying the above wave equation can thus be decomposed in terms of these modes as $\psi=\sum_{m,n}C_{m,n} u_{m, n}$. A fundamental question then arises: How can one measure the mode spectrum (weights) $\mathcal{P}_{m,n}=\vert C_{m,n}\vert^{2}$ of any given scalar field?
\par
To answer the above question we apply a symplectic invariant approach~\cite{Simon00,Calvo06,Calvo08}. Symplectic methods have been used in theories of elementary particles, condensed matter, accelerator and plasma physics, oceanographic and atmospheric sciences, and optics~\cite{Guillemin}. Central to our work is the recognition that any linear {\em passive} symplectic transformation $S$ acting on the canonical Hermitian operators $\hat{x},\hat{y},\hat{p}_{x}$, and $\hat{p}_{y}$ (whose only  nonvanishing commutators are $[\hat{x},\hat{p}_{x}]=[\hat{y},\hat{p}_{y}]=i\lambdabar$), is associated with a unitary operator $\hat{U}(S)$ generated by the group
\begin{eqnarray}
\hat{\mathcal{L}}_{x}\!\!&=&\!\! \frac{\hat{x}^{2}-\hat{y}^{2}}{2w_{0}^{2}}+\frac{(\hat{p}_{x}^{2}-\hat{p}_{y}^{2})w_{0}^{2}}{8\lambdabar^{2}}\, , \quad \hat{\mathcal{L}}_{y}= \frac{\hat{x}\hat{y}}{w_{0}^{2}}+\frac{\hat{p}_{x}\hat{p}_{y}w_{0}^{2}}{4\lambdabar^{2}}\, , \nonumber\\
\hat{\mathcal{L}}_{z} \!\!&= &\!\! \frac{\hat{x}\hat{p}_{y}-\hat{y}\hat{p}_{x}}{2\lambdabar}\, ,\quad \hat{\mathcal{N}}= \frac{\hat{x}^{2}+\hat{y}^{2}}{2w_{0}^{2}}+\frac{(\hat{p}_{x}^{2}+\hat{p}_{y}^{2})w_{0}^{2}}{8\lambdabar^{2}}\, . 
\label{eq:passiveoperators}
\end{eqnarray}
Here, $\lambdabar=1/k$ and $w_{0}$ is the characteristic width of the spatial modes $u_{m,n}$ in which the analyzed wave field is to be decomposed. The set~(\ref{eq:passiveoperators}) satisfies the usual SU$(2)$ algebra $[\hat{\mathcal{L}}_{a},\hat{\mathcal{L}}_{b}]=i\varepsilon_{abc}\hat{\mathcal{L}}_{c}$ ($a,b,c=x,y,z$), with $\hat{\mathcal{N}}$ being the only commuting generator in the group and $\hat{\mathcal{L}}_{z}$ describing real spatial rotations on the transverse $xy$ plane (it is proportional to the orbital angular momentum operator along the wave propagation direction~\cite{Calvo06}). We can therefore represent the most general passive unitary operator $\hat{U}(S)$ associated with $S$ by a single exponential of linear combinations of any of the above generators~\cite{Calvo08} $\hat{U}(S)=\exp[-i(\phi_{+}\,\hat{\mathcal{N}}+{\boldsymbol\Phi}_{-}\cdot\hat{\boldsymbol{\mathcal{L}}})]$, with real scalar $\phi_{+}$ and vector ${\boldsymbol\Phi}_{-}$ parameters. We show below how $\hat{U}(S)$ can be implemented using simple optical elements.
\par
The method proposed here relies on exploiting $\hat{\mathcal{N}}$, together with specific combinations of the generators $\hat{\mathcal{L}}_{x}$ and $\hat{\mathcal{L}}_{y}$, to construct two commuting unitary operators from which the associated symplectic matrices and their experimental implementation can be found relatively easily. To this end, let $\vert m, n\rangle$ denote the (pure) mode states of order $m+n\geq0$ (in position-representation $\langle x,y\vert m, n\rangle=u_{m,n}$). We first impose $\vert m, n\rangle$ to be eigenstates of both $\hat{\mathcal{N}}$ and the operator $\hat{\mathcal{L}}_{\theta,\varphi}\equiv {\bf u}_{r}\cdot\hat{\boldsymbol{\mathcal{L}}}$, where ${\bf u}_{r}=(\cos\varphi\sin\theta,\sin\varphi\sin\theta,\cos\theta)$ may be conceived as a radially oriented unit vector in the orbital Poincar\'{e} sphere~\cite{Padgett99,Calvo05}. The eigenstates $\vert m, n\rangle$ depend on the choice of $\theta$ and $\varphi$ and fulfill $\hat{\mathcal{N}}\vert m, n\rangle=[(m+n)/2]\vert m, n\rangle$ and $\hat{\mathcal{L}}_{\theta,\varphi}\vert m, n\rangle=[(m-n)/2]\vert m, n\rangle$.  For instance, the eigenvectors of operators $\hat{\mathcal{L}}_{x}$ and $\hat{\mathcal{L}}_{z}$ are the HG and LG modes, respectively~\cite{Calvo06,Calvo05}. Measurements of $\vert \psi\rangle=\sum_{m,n}C_{m,n}\vert m, n\rangle$ will involve the action of the unitaries $\hat{U}_{\mathcal{N}}=e^{-i\phi_{+}\,\hat{\mathcal{N}}}$ and $\hat{U}_{\theta,\varphi}=e^{-i\phi_{-}\hat{\mathcal{L}}_{\theta,\varphi}}$, upon variation of parameters $\phi_{+}$ and  $\phi_{-}$, which can be externally controlled. Notice that $\hat{U}_{\mathcal{N}}$ is connected with the Gouy phase~\cite{Gouy}, whereas  $\hat{U}_{\theta,\varphi}$ describes $\phi_{-}$-angle rotations about ${\bf u}_{r}$, thereby changing the mode superpositions. 
\par
In order to extract the complete spectrum $\mathcal{P}_{m,n}$ of a coherent electromagnetic scalar field, an optical scheme is further developed.  We remark that for quantum mechanical matter waves (e.g. Bose-Einstein condensates), a conceptually similar approach to analyze their spatial structure should currently be feasible by exploiting atom optics: interferometers~\cite{AtomInterf}, beam splitters in atom chips~\cite{Cassettari}, focusing and storage in resonators~\cite{Bloch}, and conical lenses~\cite{Muniz}.  We use here a Mach-Zehnder interferometer with built-in refractive components performing $\hat{U}_{\mathcal{N}}$ and $\hat{U}_{\theta,\varphi}$: sets of spherical and cylindrical thin lenses. For ease of operation, it is desirable to vary  $\phi_{\pm}$ in a way that minimizes the displacements of the optical elements. We explicitly show that all the required transformations in the interferometer can be achieved solely by rotation and variation of the focal lengths of the lenses. Translations of the lenses are not necessary, although it may turn out to be more convenient in certain instances to carry finite displacements in some of them. In any case, the arms of the interferometer always remain fixed.
\par
Detection of the light intensity difference $\Delta I\equiv I_{B}-I_{A}$ at the two output ports (A and B) of the interferometer provides the data for reconstructing the weights $\mathcal{P}_{m,n}$. One has $\Delta I\propto\langle\psi\vert (\hat{U}^{\dagger}_{\mathcal{N}}\hat{U}^{\dagger}_{\theta,\varphi} \hat{U}_{C}+\hat{U}^{\dagger}_{C}\hat{U}_{\mathcal{N}}\hat{U}_{\theta,\varphi})\vert\psi\rangle$, where $\hat{U}_{C}$ represents the unitary operator for the compensating lens system in the complementary arm of the interferometer. For clarity, let us assume that $\hat{U}_{C}$ comprises a similar lens set as the one for $\hat{U}_{\mathcal{N}}\hat{U}_{\theta,\varphi}$ (we will show later that this assumption can be removed), with equivalent parameters $\phi_{+}'$ and  $\phi_{-}'$. The mode spectrum and the output intensity difference are directly connected by a double Fourier-like transform
\begin{eqnarray} 
\mathcal{P}_{m,n}&\propto& \int_{0}^{4\pi}\!\!\int_{0}^{4\pi} d\phi_{+}\,d\phi_{-}\,\Delta I(\phi_{+}-\phi_{+}', \phi_{-}-\phi_{-}') \nonumber\\
&\times& e^{i(m+n)(\phi_{+}-\phi_{+}')/2} \, e^{i(m-n)(\phi_{-}-\phi_{-}')/2} , 
\label{eq:Spectrum}
\end{eqnarray}
where the proportionality constant equals $(16\pi^{2})^{-1}$ if $m=n=0$, and  $(8\pi^{2})^{-1}$ otherwise. Note that $\Delta I$ possess the symmetry properties $\Delta I(\phi_{+}\pm2\pi, \phi_{-}\pm2\pi)=\Delta I(\phi_{+}\pm2\pi, \phi_{-}\mp2\pi)=\Delta I(\phi_{+}, \phi_{-})$. Our scheme nontrivially generalizes that of Ref.~\cite{Zambrini}, aimed to reveal the quasi-intrinsic nature of the orbital angular momentum degree of freedom for scalar waves. There, by means of the measurement $\hat{U}= e^{-i\phi \hat{\mathcal{L}}_{z}}$, implemented with Dove prisms, the azimuthal index spectrum of spiral harmonic modes could readily be accessed. Here, we explore the entire transverse mode space expanded by the indices $m$ and $n$. Hence, the combined effect of the two nonequivalent measurements $\hat{U}_{\mathcal{N}}$ and $\hat{U}_{\theta,\varphi}$ is crucial, and would allow us to measure, for instance, either the full spectrum of HG  or LG modes.
\par
To show an explicit application of the above approach, we proceed with the characterization and design of an optical system that enables the reconstruction of the HG mode spectrum $\mathcal{P}_{n_{x},n_{y}}$ of a light beam ($n_{x},n_{y}\geq0$). In this scenario, one needs to consider the unitary operators $\hat{U}_{\mathcal{N}} $ and $\hat{U}_{\theta=\pi/2,\phi=0} = e^{-i\phi_{-} \hat{\mathcal{L}}_{x}} $. The path from these unitaries to the pursued setups resorts to the Stone-von Neumann theorem~\cite{Simon00,Calvo06,Calvo08}; it yields the symplectic matrices $S_{+}$ and $S_{-}$ which contain all systems information
\begin{eqnarray}
S_{\pm} = \left( \begin{array}{cccc}
c_{\pm} & 0& z_{0}s_{\pm}& 0 \\
0&c_{\pm} & 0&\pm z_{0}s_{\pm} \\
-s_{\pm}/z_{0} & 0 & c_{\pm} & 0\\
0&\mp s_{\pm}/z_{0} & 0 & c_{\pm}
\end{array} \right)\! ,
\label{eq:Spm}
\end{eqnarray}
where $c_{\pm}=\cos(\phi_{\pm} /2)$, $s_{\pm}=\sin(\phi_{\pm} /2)$, and $z_{0}=w_{0}^{2}/(2\lambdabar)$ is the Rayleigh range. From matrices $S_{\pm}$ one can then obtain the corresponding integral transforms that govern the propagation (along the $z$-direction) of any input paraxial wave $\psi(x,y)$ traversing each system. These transforms lead, in our case, to the integral kernels
\begin{eqnarray}
K_{\pm}(x,y;x',y')\!&=&\!\frac{1}{\pi i\vert s_{\pm}\vert w_{0}^{2}}\exp\left[\frac{-2i(xx'\pm yy')}{w_{0}^{2}s_{\pm}}\right]\nonumber\\
&\times&\!\!\exp\left[\frac{i(x^{2}\pm y^{2}+x^{'2}\pm y^{'2})c_{\pm}}{w_{0}^{2}s_{\pm}}\right]\! .
\label{eq:Kernels}
\end{eqnarray}
The output wave profiles after each $S_{\pm}$ result from $\psi_{\pm}(x,y)=\int dx' dy' K_{\pm}(x,y;x',y')\psi(x',y')$. Kernels~(\ref{eq:Kernels}) exhibit the structure of those for the fractional Fourier transform~\cite{Ozaktas}. Integral transforms associated to unitary operators generated by $\hat{\mathcal{L}}_{y}$ (rather than $\hat{\mathcal{L}}_{x}$) have been formulated~\cite{Rodrigo}. They make possible the full meridian-rotation on the orbital Poincar\'{e} sphere and, in particular, the reciprocal conversion between HG and LG modes.

\begin{figure}
\begin{center}
\hspace*{-0.0cm}
\includegraphics[width=80mm]{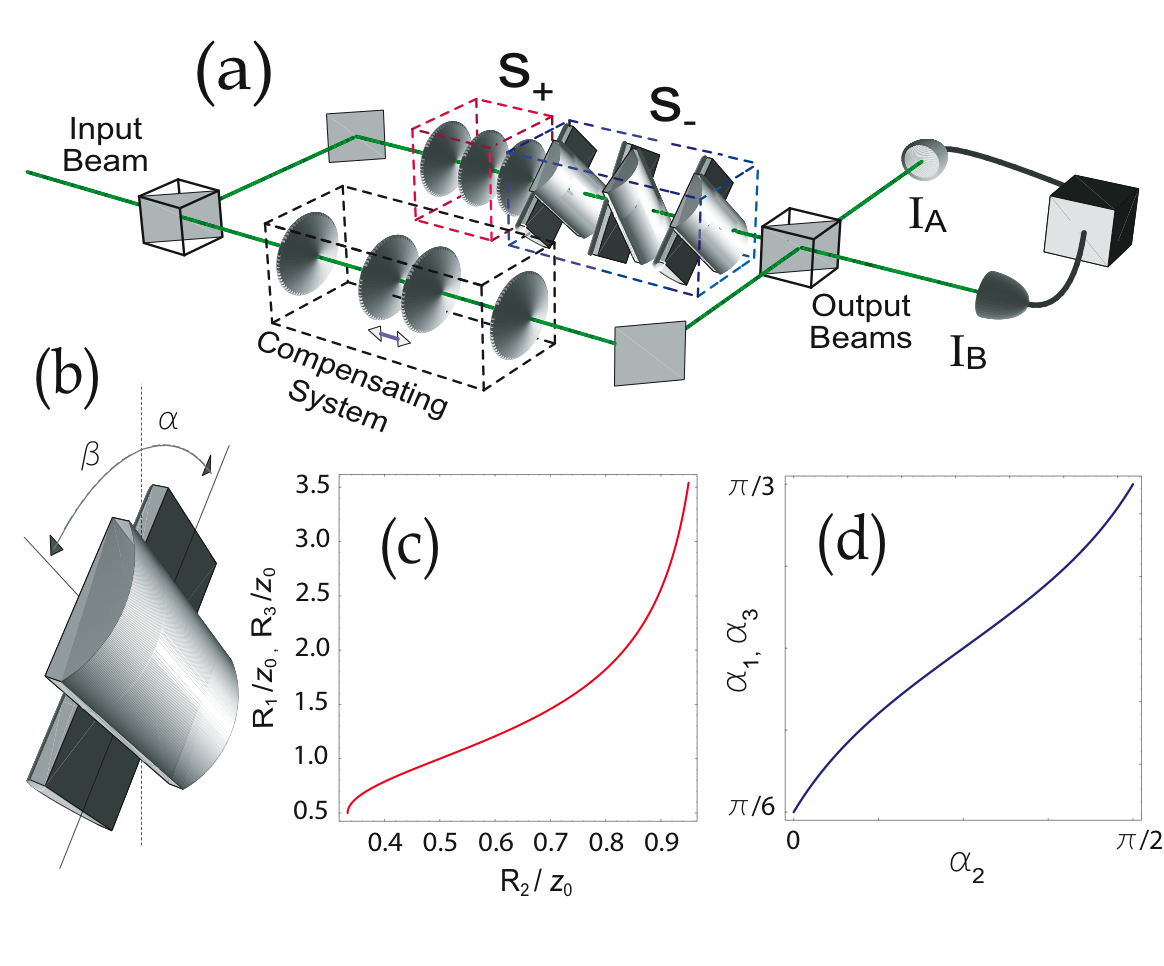}
\end{center}
\vspace*{-1cm}
\caption{\small (Color online) (a) Optical scheme for measuring the HG mode spectrum. Transformations $S_{+}$ and $S_{-}$ consist of symmetric sets with three fixed tunable-focus spherical lenses and three pairs of rotary cylindrical lenses, respectively. The compensating system involves two settings with four and two spherical lenses, respectively. (b) Detail of one of the assembled pairs of cylindrical lenses required for $S_{-}$. (c) and (d) Operation curves for $S_{+}$ and $S_{-}$, respectively (see text).}
\label{fig:Setup}
\end{figure}
\par
The two proposed optical systems for $S_{\pm}$ are symmetric and illustrated in Fig.~\ref{fig:Setup}(a). For $S_{+}$, three fixed spherical lenses with varying radii of curvature $R_{j}$ (linked to the focal lengths $f_{j}$ by $R_{j}=(\tilde{n}_{j}-1)f_{j}$, $\tilde{n}_{j}$ being the refractive indexes), and placed at equal distances $z_{0}$, are sufficient. They fulfill $R_{1}=R_{3}=z_{0}/[1-\cot(\phi_{+}/4)]$ and $R_{2}=z_{0}/(2-s_{+})$.  Tunable-focus liquid crystal spherical lenses controlled by externally applied voltages have been demonstrated, displaying  a wide range of focal lengths~\cite{Ren}. Figure~\ref{fig:Setup}(c) shows the operation curve describing the dependence between $R_{1},R_{3}$ and $R_{2}$ needed to cover the interval $\pi\leq\phi_{+}<3\pi$. These values can be attained with the lenses of Ref.~\cite{Ren}. For $S_{-}$, three pairs of cylindrical lenses are required [see Fig.~\ref{fig:Setup}(a)]. Each pair of assembled cylindrical lenses is rotated in a {\em scissor} fashion with $\alpha=-\beta$ [Fig.~\ref{fig:Setup}(b)]. The corresponding angles satisfy $\alpha_{1}=\alpha_{3}=(\pi-\Omega)/4$ and $\alpha_{2}=(3\pi-\phi_{-})/4$. The operation curve in Fig.~\ref{fig:Setup}(d) represents the variation of the rotation angles in accordance with the constraint imposed by $\cot(\phi_{-}/4)=-2\sin(\Omega/2)$.  This equation is satisfied when $\pi\leq\phi_{-}\leq3\pi$ and $-(\pi/3)\leq\Omega\leq(\pi/3)$, which lead to $0\leq\alpha_{2}\leq(\pi/2)$ and $(\pi/6)\leq\alpha_{1}\leq(\pi/3)$. The radii of curvature of the cylindrical lenses are $R_{1}=R_{2}=R_{5}=R_{6}=z_{0}/2$ and $R_{3}=R_{4}=z_{0}/4$. The distance between consecutive pairs is $z_{0}/2$. With this scheme, the covered values for $\phi_{-}\in[\pi,3\pi]$. The fact that both $\phi_{+}$ and $\phi_{-}$ are restricted to the interval $[\pi,3\pi]$, rather than to $[0,4\pi]$, does not constitute a fundamental limitation. It can be circumvented by employing the compensating system [see Fig.~\ref{fig:Setup}(a)] and the symmetry properties of $\Delta I$. If two sequences of measurements are made, each having a different compensating system performing transformations with $\phi_{+}'=\phi_{-}'=0$ (identity matrix) and $\phi_{+}'=0,\phi_{-}'=2\pi$ (minus identity matrix), respectively, then Eq.~(\ref{eq:Spectrum}) can be cast ($m\to n_{x}$ and $n\to n_{y}$) as
\begin{eqnarray} 
\mathcal{P}_{n_{x},n_{y}}\!\!&\propto&\!\!\int_{\pi}^{3\pi}\!\!\int_{\pi}^{3\pi}\! d\phi_{+}\,d\phi_{-}\!\left[(-1)^{n_{x}+n_{y}}\Delta I(\phi_{+}, \phi_{-}-2\pi)\right.\nonumber\\
&+&\!\!\left.\Delta I(\phi_{+}, \phi_{-})\right]\!\cos\!\left[\frac{(n_{x}+n_{y})\phi_{+}+(n_{x}-n_{y})\phi_{-}}{2}\right]\!.\nonumber\\
\label{eq:SpectrumHG}
\end{eqnarray}
The proportionality constant is $(8\pi^{2})^{-1}$ if $n_{x}=n_{y}=0$, and  $(4\pi^{2})^{-1}$ otherwise. The integration intervals in~(\ref{eq:SpectrumHG}) now display the accessible ranges for  $\phi_{\pm}$. The first and second sequences of measurements can be carried out with a compensating system made of four and two (all identical) spherical lenses, respectively. By properly choosing their focal lengths, it is not necessary to displace the $S_{\pm}$ systems nor the interferometer arms. Moreover, measuring the LG spectra would only require embedding the system $S_{-}$ between two fixed, mutually orthogonal, cylindrical lenses, with $S_{+}$ remaining invariant.
\par
\begin{figure}
\begin{center}
\hspace*{0.0cm}
\hbox{\vbox{\vskip -0.0cm \includegraphics[width=90mm]{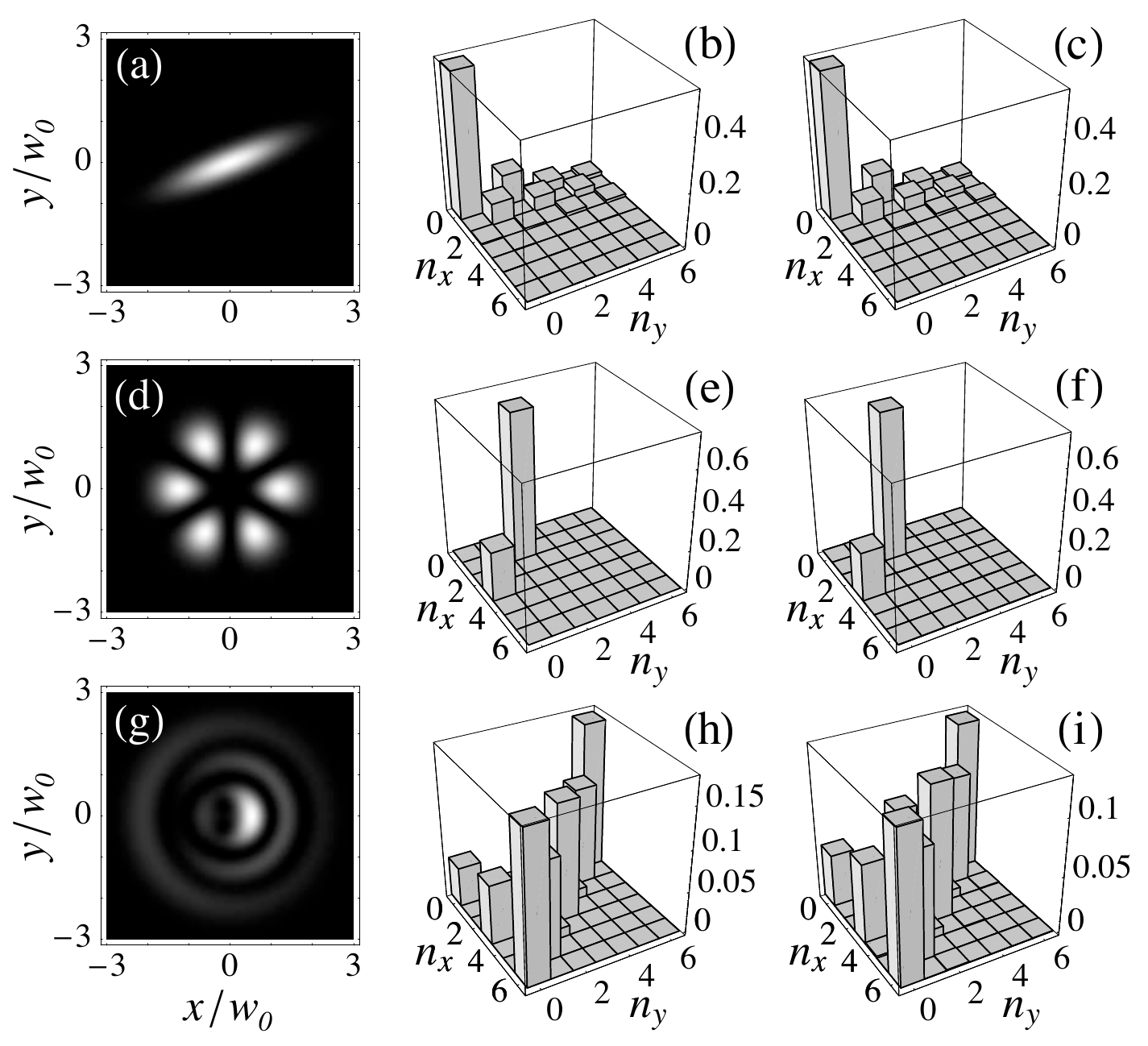}}}
\end{center}
\vspace*{-0.4cm}
\caption{\small Mode spectrum analysis. Exact and reconstructed [via Eq.~(\ref{eq:SpectrumHG})] HG mode weights (central and right columns, respectively) from the input beam profiles (left column).}
\label{fig:Simulation}
\end{figure}

To demonstrate that our data analysis scheme is feasible and does not demand a large number of measurements for each $\phi_{\pm}$, we have numerically simulated the transformation and processing of several input coherent waves. Figure~\ref{fig:Simulation} depicts the light profiles entering into the interferometer: strongly astigmatic Gaussian [Fig.~\ref{fig:Simulation}(a)], hexapole necklace [Fig.~\ref{fig:Simulation}(d)] and anisotropic multiring [Fig.~\ref{fig:Simulation}(g)] beams. Their exact HG distributions together with those retrieved from Eq.~(\ref{eq:SpectrumHG}) are also displayed. To simulate the evolution of the various beams, the kernels~(\ref{eq:Kernels}) were used to calculate $\Delta I$ in Eq.~(\ref{eq:SpectrumHG}) for $10$ different values (in equal increments) per each $\phi_{\pm}\in[\pi,3\pi]$. The reconstructed and exact weights agree well (compare right and central columns in Fig.~\ref{fig:Simulation}). In analogy with the Whittaker-Shannon sampling theorem in Fourier analysis~\cite{Goodman}, input beams that are {\em mode-band-limited} can be {\em exactly} reconstructed via our scheme [compare Figs.~\ref{fig:Simulation}(e) and \ref{fig:Simulation}(f)].  That is, if the sampling increments for $\phi_{\pm}$ are smaller than the inverse of the highest contributing mode numbers, reconstruction will be exact. Misalignment of the optical elements is expected to be the main source of errors. The effect on the weights $\mathcal{P}_{n_{x},n_{y}}$, due to lens displacements (tolerances) $\delta$ with respect to the beam axis, introduces a correction term $\sim (\delta/w_{0})^{2}$, which is smaller than $1\%$ for typical values $\delta\lesssim 10\mu$m and $w_{0}\geq100\mu$m.
\par
In conclusion, we have proposed a spatial spectrum analyzer for waves with arbitrary profiles based on a simple setup of lenses in a Mach-Zehnder interferometer. Measuring the mode spectrum of
optical beams is fundamental not only to achieve control of laser beam profiles but also to access information encoded in this degree of freedom of the electromagnetic field, in the context of spatial multiplexing. The generality of the presented theoretical analysis encompasses different mode bases and suggests the measurement strategy also for matter waves. The specific set-up for HG spectra is presented with details in view of a more immediate experimental realization. Moreover, our scheme is compatible with mode analysis at the single-photon level, and could be of great use for quantum information applications. Retrieval of the complete spatial mode spectrum of any monochromatic optical field would become an attractive tool for signal analysis and processing when combined with novel holographic recording materials designed for modal tailoring and multiplexors. In this respect, amorphous photopolymerizable glasses, exhibiting high refractive index modulation in experimental demonstrations of the optical Pendell\"{o}sung effect~\cite{Pendellosung},  are well suited for this purpose.
\par
We acknowledge financial support  from the Spanish Ministry of Science and Technology through Projects FIS2005-01369 and FIS2006-04190, from Govern Balear (PROGECIB-5A), Junta de Comunidades de Castilla-La Mancha (PCI-08-0093), Juan de la Cierva and Ramon y Cajal Grant Programs, and Consolider Ingenio 2010 QIOT CSD2006-00019.
\par
*email address: Gabriel.Fernandez@uclm.es 
\par

\end{document}